# Modeling and experimental research of the processes of radiation damage formation in the interaction of gamma quanta flows and relativistic electron beams with solution of organic dyes

*S.P. Gokov[1], Yu.G. Kazarinov[1,2], S.A. Kalenik[1], V.Y. Kasilov[1], V.V. Kantemirov[1], O.O. Mazilov[1], T.V. Malykhina[1,2], V.V. Tsiats'ko[1], E.V. Tsiats'ko[1]*

[1]*National Science Center "Kharkov Institute of Physics and Technology", Kharkiv, Ukraine;*
[2]*V.N. Karazin Kharkiv National University, Kharkiv, Ukraine*
*E-mail: stek@kipt.kharkov.ua*

The processes of interaction of an aqueous solution of methylene blue (MB) organic dye ($C_{16}H_{18}N_3SCl$) with gamma quanta and electrons have been investigated. A model has been developed, and the passage of electrons with an energy of 15 MeV through tungsten layers with a thickness of 1…8 mm has been simulated. A number of experiments were carried out on the basis of the simulation results. Analysis of the calculated and experimental data showed that one incident electron destroys 4 times more dye molecules than one incident gamma-quantum.

## INTRODUCTION

Investigation of the processes of interaction of ionizing radiation with complex organic objects makes it possible to solve a number of applied and fundamental problems in the field of radiation physics, chemistry and biology. These tasks include: radiation-chemical protection, the study and increase of the resistance of organic materials and biological systems to the action of ionizing radiation, the development of new radioprotectors and new dosimeters, the purposeful change of properties and radiation-chemical synthesis of materials, as well as the use of ionizing radiation for medical purposes [1, 2].

The goal of this work was to research the processes occurring during the interaction of an aqueous solution of an organic dye: methylene blue (MB) – $C_{16}H_{18}N_3SCl$ with gamma-quanta and high-energy electrons. The work was carried out on a LINAC LUE-300 at NSC KIPT. The use of aqueous solutions of organic dyes makes it possible to study processes similar to those that occur during the interaction of ionizing radiation with biological objects. This is very important for research in nuclear medicine and biology. This technique also makes it possible to analyze the dynamics of destruction of an organic object at different absorbed doses, which makes it possible to use them as dosimeters of ionizing radiation.

## MODEL OF EXPERIMENT

To irradiate organic dye solutions with gamma-quanta and high-energy electrons, a tungsten bremsstrahlung converter was used at the LUE-300 output. The task of this work was also to determine the dependence of the intensity of radiation destruction of the organic dye solution on the thickness of the tungsten bremsstrahlung converter. The transverse dimensions of the tungsten plate were 50×50 mm. Energy of primary electrons was 15 MeV for all presented experiments.

To determine the dependence of the intensity of the radiation destruction of the organic dye solution on the thickness of the converter, a model was developed and a simulation of the passage of an electron beam with a primary energy of 15 MeV through tungsten layers with a thickness of 1 to 8 mm was carried out. The carrier of the model is a computer program developed in the C++ language using the Geant4 class library. The simulation based on the Geant4 library uses the Monte Carlo method.

To obtain the minimum statistical error of calculations, a large number of events are usually used, but at the same time the calculation time can be significant, and the arrays of the processed data can be excessively large. Therefore, in order to minimize the statistical error of calculations, and also taking into account the need for further processing of the results, the number of primary particles was chosen $N_e=10^7$. Such a number of events will allow for a finite time to make the necessary calculations with a statistical error of less than 1%, which fully satisfies the task. During the calculations, the "emstandard_opt3" model of the "PhysicsList" module was selected, since this model is the most suitable model [3, 4] and adequately describes all the necessary processes in the considered energy range for all primary and secondary particles.

The block for describing the geometric parameters of the simulated setup corresponded to the geometric parameters of the real experimental setup shown in Fig. 1.

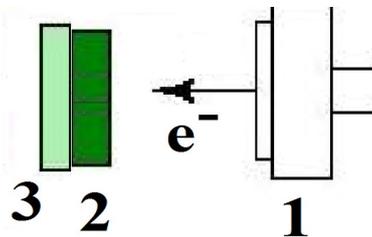

*Fig. 1. Schematic of the experimental setup:*
*1 – output of LINAC; 2 – tungsten converter; 3 – target*

The tungsten converter was placed at a distance of 50 mm from the titanium foil of the exit window of the linear electron accelerator. In Fig. 2 shows the energy spectra of bremsstrahlung gamma radiation emitted forward from a 50×50 mm tungsten transducer for different thicknesses of the tungsten plate – from 1 to 5 mm with a step of 1 mm. The results are normalized to 1 primary electron. The graph (see Fig. 2) is presented on a logarithmic scale along the X and Y axes for ease of comparison of energy spectra. It can be seen what increasing of thickness of converter mainly unaffect on gamma spectra at the energy more than 400 keV and leads to a significant decrease of intensity below this energy.

It can be noted that the maximum yield of bremsstrahlung gamma quanta is observed at a tungsten plate thickness of 2 mm, and slightly decreases at a plate thickness of 3 mm. However, the most valuable information will be information on the amount of inhibitory gamma-quanta immediately in front of the target containing an organic dye solution.

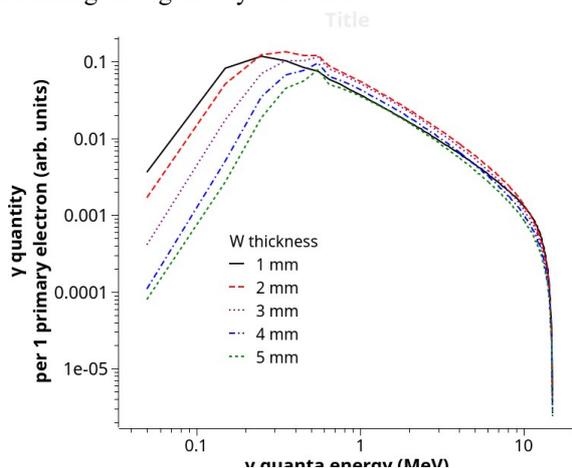

*Fig. 2. Energy spectra of bremsstrahlung gamma-quanta emitted by the W-converter for different tungsten thickness – from 1 to 5 mm*

The flux of bremsstrahlung gamma-quanta and the flux of electrons immediately in front of the target containing the dye solution were estimated. The calculation results for various thicknesses W of the converter are shown in Fig. 3.

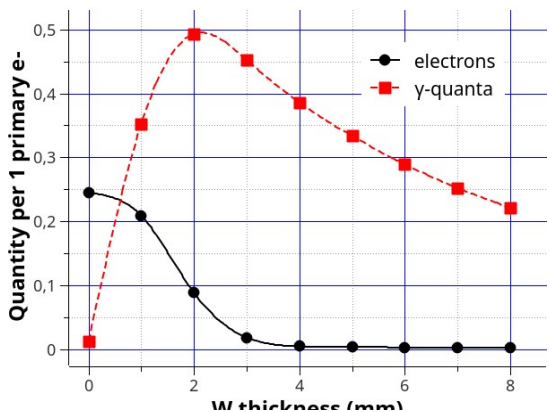

*Fig. 3. The flux of bremsstrahlung gamma-quanta and electrons directly in front of the target (for a beam with a diameter of 20 mm)*

It can be seen (see Fig. 3) that the most intense flux of bremsstrahlung gamma-quanta in front of the target is observed when the converter is 2 mm thick, but in this case there is also a significant amount of electrons in front of the target. Therefore, for further calculations, a 4 mm thick converter was chosen, at which the flux of bremsstrahlung gamma-quanta in front of the target decreased slightly (see Fig. 3), but the electron flux is negligible and consists mainly of low-energy electrons.

The ratio of the fluxes of electrons and gamma-quanta immediately in front of the target was also obtained (Fig. 4).

Simulation of the passage of an electron beam through a target containing a dye solution was carried out in order to determine the contribution of various processes to the total value of the energy absorbed in the target. The results are shown in Fig. 5.

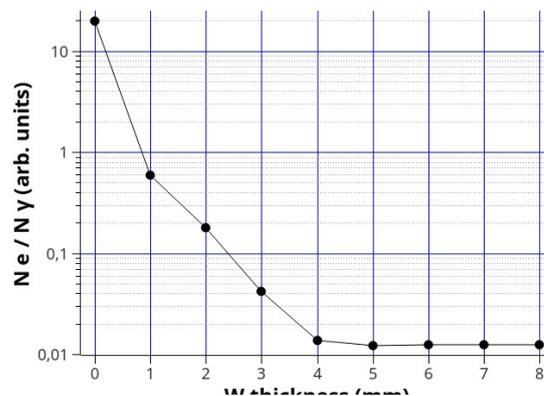

*Fig. 4. Ratio of fluxes of electrons and gamma-quanta directly in front of the target*

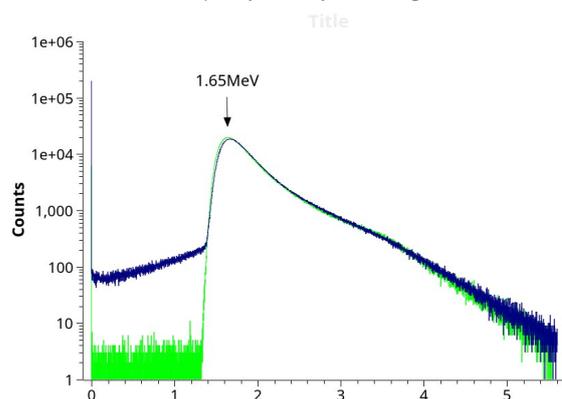

*Fig. 5. Energy absorbed in the target (green curve – only ionization energy losses are taken into account, dark blue curve – all processes are taken into account)*

## EXPERIMENTAL PART

The experiment was carried out on a LINAC LUE-300. The electron beam energy was 15 MeV at a current density of 1 μA/cm$^2$. Irradiation was carried out both directly with an electron beam and using a converter located in front of the target. The converter consisted of tungsten plates with a total thickness of 2, 4, and 6 mm. The target was 3 cm$^3$ dye solution placed on the path of the electron beam in a rigidly fixed glass tube. The use of a fixed stand made it possible to set the tube in a constant position when changing solutions relative to the exit window of the accelerator.

A photo of the experimental setup is shown in Fig. 6.

The absorption spectra of MB aqueous solutions before and after irradiation are shown in Fig. 7. The spectra were measured on an SF-46 single-beam spectrophotometer in the range of the main absorption peak of the dye from 500 to 800 nm. To reduce the error in measuring the absorption, the concentration of the dye in the initial solution was selected so that the absorption at the absorption peak was near 1.

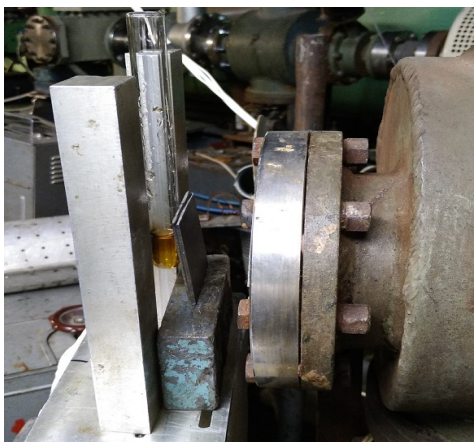

*Fig. 6. Experimental setup*

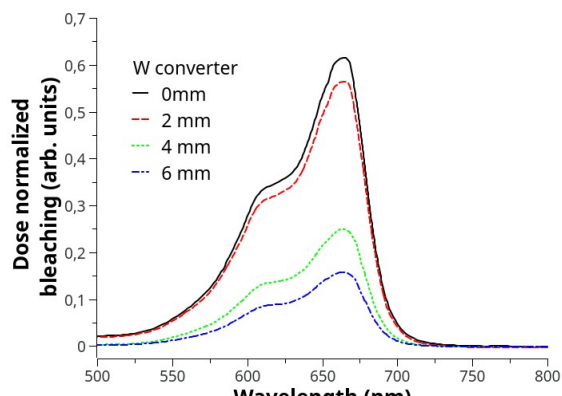

*Fig. 8. Radiation bleaching of organic dye solution normalized to 1 s irradiation with different converter thicknesses*

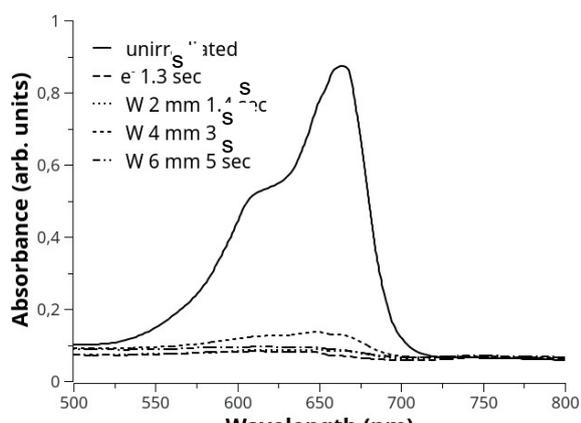

*Fig. 7. Absorption spectra of a solution of an organic dye methylene blue before and after irradiation with a tungsten converter of various thicknesses*

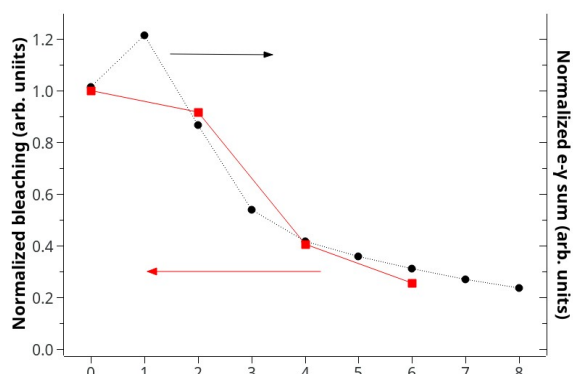

*Fig. 9. Normalized radiation bleaching of the organic dye solution (squares) and weighted sum of the calculated number of electrons and gamma-quanta (circles) depending on the thickness of the tungsten converter*

As can be seen, even short-term (near 1s.) irradiation leads to significant degradation of the dye in solution. From Fig. 8, which revealed the differential absorption spectra (before and after irradiation) normalized to the unit of irradiation time (dose), a decreasing of level of discoloration at increasing the thickness of the tungsten converter is observed.

In Fig. 9 shows the normalized dependence of the absorption at a wavelength of 685 nm of an aqueous solution of the dye on the thickness of the tungsten converter and the weighted sum of the calculated number of electrons and gamma-quanta that entered the solution. Without taking into account the change in the spectra of electrons and gamma-quanta, the loss of color of the dye solution when exposed to gamma-quanta occurs 4 times slower than when it is exposed to electrons.

Figs. 8 and 9 shown that the greatest destruction of the MB dye occurs when irradiated with an electron beam without a converter. With the installation of the converter and the subsequent increase in its thickness, the amount of destruction of the dye molecules gradually decreases. However, the rate of decrease is much less than the decrease in the number of electrons passed through the converter, which is associated with additional irradiation in the flux of gamma-quanta.

## CONCLUSIONS

The processes occurring during the interaction of an aqueous solution of an organic dye methylene blue $C_{16}H_{18}N_3SCl$ with gamma-quanta and 15 MeV electrons have been investigated. To irradiate the MB solution with gamma-quanta, a tungsten converter of bremsstrahlung of various thicknesses was used.

A model has been developed and simulation of the passage of an electron beam with a primary energy of 15 MeV through tungsten layers of various thicknesses from 1 to 8 mm has been carried out. This simulation was carried out in order to determine the dependence of the intensity of radiation destruction of the organic dye solution on the thickness of the converter used.

The absorption spectra of an organic dye solution for a converter with a thickness of 2, 4, and 6 mm were obtained experimentally.

An analysis of the calculated and experimental data showed the greatest destruction of the dye occurs when irradiated with an electron beam, one incident electron accounts for 4 times more destruction of molecules than one incident gamma-quantum.

Moreover, with electron irradiation, 97% of the destruction of the dye occurs due to ionization, and 3% due to other processes (elastic and inelastic interaction with the nucleus, cascade formation processes, etc.).

The presented calculations and experimental results demonstrate that organic dye solutions are excellent indicators of the total absorbed dose when exposed to ionizing radiation. And they represent a suitable object for research in the field of nuclear medicine.